\documentclass{sigplan-proc}
\usepackage{epsfig}

\long\def\com#1{}

\begin{document}

\conferenceinfo{ICFP'02,} {October 4-6, 2002, Pittsburgh, Pennsylvania, USA.}
\CopyrightYear{2002}
\crdata{1-58113-487-8/02/0010}

\title{Packrat Parsing: \\
	Simple, Powerful, Lazy, Linear Time}
\subtitle{Functional Pearl}

\numberofauthors{1}
\author{
\alignauthor Bryan Ford \\
	\affaddr{Massachusetts Institute of Technology}\\
	\affaddr{Cambridge, MA}\\
	\email{baford@lcs.mit.edu}
}

\maketitle

\begin{abstract}
Packrat parsing is a novel technique for implementing parsers
in a lazy functional programming language.
A packrat parser provides the power and flexibility
of top-down parsing with backtracking and unlimited lookahead,
but nevertheless guarantees linear parse time.
Any language defined by an LL($k$) or LR($k$) grammar
can be recognized by a packrat parser,
in addition to many languages
that conventional linear-time algorithms do not support.
This additional power
simplifies the handling of common syntactic idioms
such as the widespread but troublesome longest-match rule,
enables the use of sophisticated disambiguation strategies
such as syntactic and semantic predicates,
provides better grammar composition properties,
and allows lexical analysis to be integrated seamlessly into parsing.
Yet despite its power,
packrat parsing shares the same simplicity and elegance
as recursive descent parsing;
in fact
converting a backtracking recursive descent parser
into a linear-time packrat parser
often involves only a fairly straightforward structural change.
This paper describes packrat parsing informally
with emphasis on its use in practical applications,
and explores its advantages and disadvantages
with respect to the more conventional alternatives.
\end{abstract}




\section{Introduction}

There are many ways to implement a parser
in a functional programming language.
The simplest and most direct approach
is {\em top-down} or {\em recursive descent parsing},
in which the components of a language grammar
are translated more-or-less directly
into a set of mutually recursive functions.
Top-down parsers can in turn be divided into two categories.
{\em Predictive parsers}
attempt to predict what type of language construct
to expect at a given point
by ``looking ahead'' a limited number of symbols in the input stream.
{\em Backtracking parsers}
instead make decisions speculatively
by trying different alternatives in succession:
if one alternative fails to match,
then the parser ``backtracks'' to the original input position
and tries another.
Predictive parsers are fast and guarantee linear-time parsing,
while backtracking parsers are both conceptually simpler and more powerful
but can exhibit exponential runtime.

This paper presents a top-down parsing strategy
that sidesteps the choice between prediction and backtracking.
{\em Packrat parsing}
provides the simplicity, elegance, and generality of the backtracking model,
but eliminates the risk of super-linear parse time,
by saving all intermediate parsing results as they are computed
and ensuring that no result is evaluated more than once.
The theoretical foundations of this algorithm
were worked out in the 1970s~\cite{aho72parsing,birman73parsing},
but the linear-time version was apparently never put in practice
due to the limited memory sizes of computers at that time.
However, on modern machines the storage cost of this algorithm
is reasonable for many applications.
Furthermore,
this specialized form of memoization
can be implemented very elegantly and efficiently
in modern lazy functional programming languages,
requiring no hash tables or other explicit lookup structures.
This marriage of a classic but neglected linear-time parsing algorithm
with modern functional programming
is the primary technical contribution of this paper.

Packrat parsing is unusually powerful
despite its linear time guarantee.
A packrat parser can easily be constructed
for any language described by an LL($k$) or LR($k$) grammar,
as well as for many languages that require unlimited lookahead
and therefore are not LR.
This flexibility eliminates many of the troublesome restrictions
imposed by parser generators of the YACC lineage.
Packrat parsers are also much simpler to construct than bottom-up LR parsers,
making it practical to build them by hand.
This paper explores the manual construction approach,
although automatic construction of packrat parsers
is a promising direction for future work.

A packrat parser can directly and efficiently implement
common disambiguation rules
such as {\em longest-match}, {\em followed-by}, and {\em not-followed-by},
which are difficult to express unambiguously in a context-free grammar
or implement in conventional linear-time parsers.
For example, recognizing identifiers or numbers during lexical analysis,
parsing {\tt if}-{\tt then}-{\tt else} statements in C-like languages,
and handling {\tt do}, {\tt let}, and lambda expressions in Haskell
inherently involve longest-match disambiguation.
Packrat parsers are also
more easily and naturally composable than LR parsers,
making them a more suitable substrate for dynamic or extensible syntax%
~\cite{adams91modular}.
Finally, both lexical and hierarchical analysis
can be seamlessly integrated into a single unified packrat parser,
and lexical and hierarchical
language features can even be blended together,
so as to handle string literals with embedded expressions
or literate comments with structured document markup,
for example.

The main disadvantage of packrat parsing is its space consumption.
Although its asymptotic worst-case bound
is the same as those of conventional algorithms---%
linear in the size of the input---%
its space utilization is directly proportional to input size
rather than maximum recursion depth,
which may differ by orders of magnitude.
However, for many applications such as modern optimizing compilers,
the storage cost of a pacrkat parser is likely to be no greater
than the cost of subsequent processing stages.
This cost may therefore be a reasonable tradeoff
for the power and flexibility of linear-time parsing with unlimited lookahead.

The rest of this paper
explores packrat parsing
with the aim of providing
a pragmatic sense of how to implement it
and when it is useful.
Basic familiarity with context-free grammars and top-down parsing is assumed.
For brevity and clarity of presentation,
only small excerpts of example code are included in the text.
However, all of the examples described in this paper are available,
as complete and working Haskell code,
at:

\begin{quote}
\begin{small}
\verb|http://pdos.lcs.mit.edu/~baford/packrat/icfp02|
\end{small}
\end{quote}

The paper is organized as follows.
Section~\ref{algorithm} introduces packrat parsing and describes how it works,
using conventional recursive descent parsing as a starting point.
Section~\ref{extensions} presents useful extensions to the basic algorithm,
such as support for left recursion, lexical analysis,
and monadic parsing.
Section~\ref{comparison} explores in more detail
the recognition power of packrat parsers
in comparison with conventional linear-time parsers.
Section~\ref{issues} discusses the three main practical limitations
of packrat parsing: determinism, statelessness, and space consumption.
Section~\ref{results} presents some experimental results
to demonstrate the practicality of packrat parsing for real languages.
Section~\ref{related} discusses related work,
Section~\ref{future} points out directions for future exploration,
and Section~\ref{conclusion} concludes.

\section{Building a Parser}
\label{algorithm}

Packrat parsing is essentially a top-down parsing strategy,
and as such packrat parsers are closely related
to recursive descent parsers.
For this reason,
we will first build a recursive descent parser for a trivial language
and then convert it into a packrat parser.

\subsection{Recursive Descent Parsing}
\label{recurse}

\begin{figure}
\begin{center}
\begin{tabular}{lcl}
Additive & $\leftarrow$ & Multitive `{\tt +}' Additive $|$ Multitive \\
Multitive & $\leftarrow$ & Primary `{\tt *}' Multitive $|$ Primary \\
Primary & $\leftarrow$ & `{\tt (}' Additive `{\tt )}' $|$ Decimal \\
Decimal & $\leftarrow$ & `{\tt 0}' $|$ $\dots$ $|$ `{\tt 9}' \\
\end{tabular}
\end{center}
\caption{Grammar for a trivial language}
\label{triv-grammar}
\end{figure}

Consider the standard approach
for constructing a recursive descent parser
for a grammar such as the trivial arithmetic expression language
shown in Figure~\ref{triv-grammar}.
We define four functions,
one for each of the nonterminals
on the left-hand sides of the rules.
Each function takes takes the string to be parsed,
attempts to recognize some prefix of the input string
as a derivation of the corresponding nonterminal,
and returns either a ``success'' or ``failure'' result.
On success, the function returns
the remainder of the input string
immediately following the part that was recognized,
along with some semantic value
computed from the recognized part.
Each function can recursively call itself and the other functions
in order to recognize the nonterminals
appearing on the right-hand sides of its corresponding grammar rules.

To implement this parser in Haskell,
we first need a type describing the result of a parsing function:

\begin{small}
\begin{verbatim}
data Result v = Parsed v String
              | NoParse
\end{verbatim}
\end{small}

In order to make this type generic
for different parse functions producing different kinds of semantic values,
the {\tt Result} type takes a type parameter $v$
representing the type of the associated semantic value.
A success result is built with the {\tt Parsed} constructor
and contains a semantic value (of type $v$)
and the remainder of the input text (of type {\tt String}).
A failure result is represented by the simple value {\tt NoParse}.
In this particular parser,
each of the four parse functions takes a {\tt String}
and produces a {\tt Result} with a semantic value of type {\tt Int}:

\begin{small}
\begin{verbatim}
pAdditive  :: String -> Result Int
pMultitive :: String -> Result Int
pPrimary   :: String -> Result Int
pDecimal   :: String -> Result Int
\end{verbatim}
\end{small}

The definitions of these functions have the following general structure,
directly reflecting the mutual recursion expressed
by the grammar in Figure~\ref{triv-grammar}:

\begin{small}
\begin{verbatim}
pAdditive  s = ... (calls itself and pMultitive) ...
pMultitive s = ... (calls itself and pPrimary) ...
pPrimary   s = ... (calls pAdditive and pDecimal) ...
pDecimal   s = ...
\end{verbatim}
\end{small}

For example, the {\tt pAdditive} function can be coded as follows,
using only primitive Haskell pattern matching constructs:

\begin{small}
\begin{verbatim}
-- Parse an additive-precedence expression
pAdditive :: String -> Result Int
pAdditive s = alt1 where

    -- Additive <- Multitive '+' Additive
    alt1 = case pMultitive s of
             Parsed vleft s' ->
               case s' of
                 ('+':s'') ->
                   case pAdditive s'' of
                     Parsed vright s''' ->
                       Parsed (vleft + vright) s'''
                     _ -> alt2
                 _ -> alt2
             _ -> alt2

    -- Additive <- Multitive
    alt2 = case pMultitive s of
             Parsed v s' -> Parsed v s'
             NoParse -> NoParse
\end{verbatim}
\end{small}

To compute the result of {\tt pAdditive},
we first compute the value of {\tt alt1},
representing the first alternative for this grammar rule.
This alternative in turn calls {\tt pMultitive}
to recognize a multiplicative-precedence expression.
If {\tt pMultitive} succeeds,
it returns the semantic value {\tt vleft} of that expression
and the remaining input {\tt s'} following the recognized portion of input.
We then check for a `{\tt +}' operator at position {\tt s'},
which if successful produces the string {\tt s''}
representing the remaining input after the `{\tt +}' operator.
Finally, we recursively call {\tt pAdditive} itself
to recognize another additive-precedence expression at position {\tt s''},
which if successful yields the right-hand-side result {\tt vright}
and the final remainder string {\tt s'''}.
If {\em all three} of these matches were successful,
then we return as the result of the initial call to {\tt pAdditive}
the semantic value of the addition, {\tt vleft + vright},
along with the final remainder string {\tt s'''}.
If any of these matches failed,
we fall back on {\tt alt2}, the second alternative,
which merely attempts to recognize
a single multiplicative-precedence expression
at the original input position {\tt s}
and returns that result verbatim, whether success or failure.

The other three parsing functions are constructed similarly,
in direct correspondence with the grammar.
Of course, there are easier and more concise ways
to write these parsing functions,
using an appropriate library of helper functions or combinators.
These techniques will be discussed later in Section~\ref{monadic},
but for clarity we will stick to simple pattern matching for now.

\subsection{Backtracking Versus Prediction}
\label{prediction}

The parser developed above is a {\em backtracking} parser.
If {\tt alt1} in the {\tt pAdditive} function fails, for example,
then the parser effectively ``backtracks'' to the original input position,
starting over with the original input string {\tt s}
in the second alternative {\tt alt2},
regardless of whether the first alternative failed to match
during its first, second, or third stage.
Notice that if the input {\tt s}
consists of only a single multiplicative expression,
then the {\tt pMultitive} function will be called twice on the same string:
once in the first alternative,
which will fail while trying to match a nonexistent `{\tt +}' operator,
and then again while successfully applying the second alternative.
This backtracking and redundant evaluation of parsing functions
can lead to parse times that grow exponentially
with the size of the input,
and this is the principal reason
why a ``naive'' backtracking strategy such as the one above
is never used in realistic parsers for inputs of substantial size.

The standard strategy for making top-down parsers practical
is to design them so that they can ``predict''
which of several alternative rules to apply
{\em before} actually making any recursive calls.
In this way it can be guaranteed
that parse functions are never called redundantly
and that any input can be parsed in linear time.
For example, although the grammar in Figure~\ref{triv-grammar}
is not directly suitable for a predictive parser,
it can be converted into an LL(1) grammar,
suitable for prediction with one lookahead token,
by ``left-factoring'' the Additive and Multitive nonterminals as follows:

\begin{center}
\begin{tabular}{lcl}
Additive & $\leftarrow$ & Multitive AdditiveSuffix \\
AdditiveSuffix & $\leftarrow$ & `{\tt +}' Additive $|$ $\epsilon$ \\
Multitive & $\leftarrow$ & Primary MultitiveSuffix \\
MultitiveSuffix & $\leftarrow$ & `{\tt *}' Multitive $|$ $\epsilon$ \\
\end{tabular}
\end{center}

Now the decision between the two alternatives for AdditiveSuffix
can be made before making any recursive calls
simply by checking whether the next input character is a `{\tt +}'.
However, because the prediction mechanism
only has ``raw'' input tokens (characters in this case) to work with,
and must itself operate in constant time,
the class of grammars that can be parsed predictively is very restrictive.
Care must also be taken
to keep the prediction mechanism consistent with the grammar,
which can be difficult to do manually
and highly sensitive to global properties of the language.
For example, the prediction mechanism for MultitiveSuffix
would have to be adjusted
if a higher-precedence exponentiation operator `{\tt **}'
was added to the language;
otherwise the exponentiation operator
would falsely trigger the predictor for multiplication expressions
and cause the parser to fail on valid input.

Some top-down parsers use prediction for most decisions
but fall back on full backtracking
when more flexibility is needed.
This strategy often yields
a good combination of flexibility and performance in practice,
but it still suffers the additional complexity of prediction,
and it requires the parser designer
to be intimately aware of where prediction can be used
and when backtracking is required.

\subsection{Tabular Top-Down Parsing}

As pointed out by Birman and Ullman~\cite{birman73parsing},
a backtracking top-down parser
of the kind presented in Section~\ref{recurse}
can be made to operate in linear time
without the added complexity or constraints of prediction.
The basic reason the backtracking parser can take super-linear time
is because of redundant calls to the same parse function
on the same input substring,
and these redundant calls can be eliminated through memoization.

Each parse function in the example
is dependent {\em only} on its single parameter, the input string.
Whenever a parse function makes a recursive call
to itself or to another parse function,
it always supplies either {\em the same} input string it was given
(e.g., for the call by {\tt pAdditive} to {\tt pMultitive}),
or a {\em suffix} of the original input string
(e.g., for the recursive call by {\tt pAdditive} to itself
after matching a `{\tt +}' operator).
If the input string is of length $n$,
then there are only $n+1$ distinct suffixes
that might be used in these recursive calls,
counting the original input string itself and the empty string.
Since there are only four parse functions,
there are at most $4(n+1)$ distinct intermediate results
that the parsing process might require.

We can avoid computing any of these intermediate results multiple times
by storing them in a table.
The table has one row for each of the four parse functions
and one column for each distinct position in the input string.
We fill the table
with the results of each parse function
for each input position,
starting at the {\em right} end of the input string
and working towards the left, column by column.
Within each column, we start from the bottommost cell and work upwards.
By the time we compute the result for a given cell,
the results of all would-be recursive calls in the corresponding parse function
will already have been computed and recorded elsewhere in the table;
we merely need to look up and use the appropriate results.

\begin{figure}
\centerline{\epsfig{file=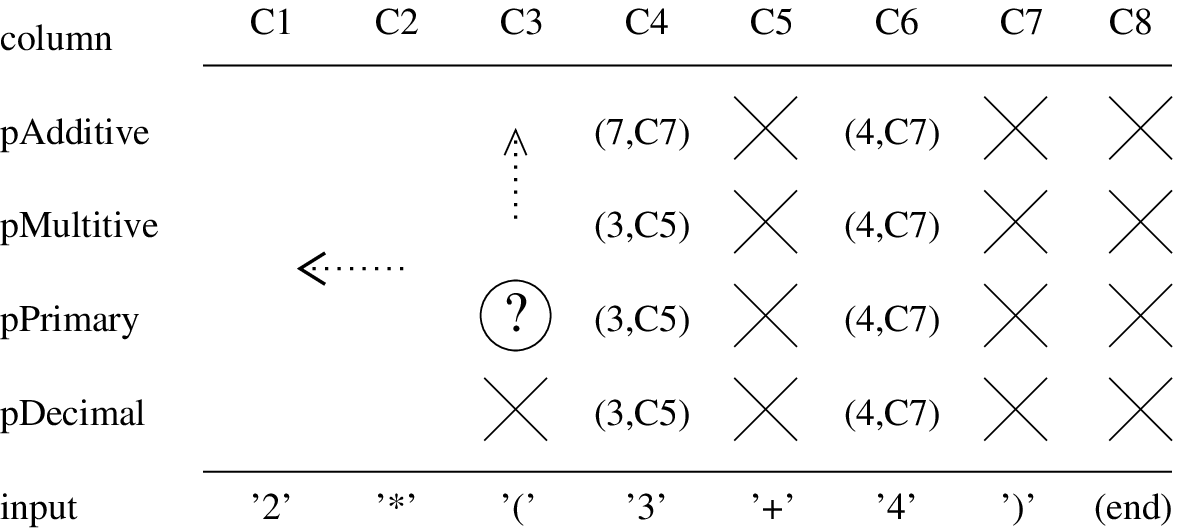, scale=0.6}}
\caption{Matrix of parsing results for string `{\tt 2*(3+4)}'}
\label{matrix}
\end{figure}

Figure~\ref{matrix} illustrates
a partially-completed result table
for the input string `{\tt 2*(3+4)}'.
For brevity, {\tt Parsed} results are indicated as {\tt ($v$,$c$)},
where $v$ is the semantic value
and $c$ is the column number
at which the associated remainder suffix begins.
Columns are labeled C1, C2, and so on,
to avoid confusion with the integer semantic values.
{\tt NoParse} results are indicated with an X in the cell.
The next cell to be filled
is the one for {\tt pPrimary} at column C3,
indicated with a circled question mark.

The rule for Primary expressions has two alternatives:
a parenthesized Additive expression or a Decimal digit.
If we try the alternatives in the order expressed in the grammar,
{\tt pPrimary} will first check for a parenthesized Additive expression.
To do so, {\tt pPrimary} first attempts to match
an opening `{\tt (}' in column C3,
which succeeds and yields as its remainder string
the input suffix starting at column C4, namely `{\tt 3+4)}'.
In the simple recursive-descent parser
{\tt pPrimary} would now recursively call {\tt pAdditive}
on this remainder string.
However, because we have the table
we can simply look up the result
for {\tt pAdditive} at column C4 in the table, which is (7,C7).
This entry indicates a semantic value of 7---%
the result of the addition expression `{\tt 3+4}'---%
and a remainder suffix of `{\tt )}' starting in column C7.
Since this match is a success,
{\tt pPrimary} finally attempts to match the closing parenthesis
at position C7,
which succeeds and yields the empty string C8 as the remainder.
The result entered for {\tt pPrimary} at column C3
is thus (7,C8).

Although for a long input string and a complex grammar
this result table may be large,
it only grows linearly with the size of the input
assuming the grammar has a fixed number of nonterminals.
Furthermore,
as long as the grammar uses only
the standard operators of Backus-Naur Form~\cite{aho86compilers},
only a fixed number of previously-recorded cells in the matrix
need to be accessed
in order to compute each new result.
Therefore, assuming table lookup occurs in constant time,
the parsing process as a whole completes in linear time.

Due to the ``forward pointers'' embedded in the results table,
the computation of a given result
may examine cells that are widely spaced in the matrix.
For example, computing the result for {\tt pPrimary} at C3 above
made use of results from columns C3, C4, and C7.
This ability to skip ahead arbitrary distances while making parsing decisions
is the source of the algorithm's unlimited lookahead capability,
and this capability makes the algorithm more powerful
than linear-time predictive parsers or LR parsers.

\subsection{Packrat Parsing}
\label{packrat}

An obvious practical problem
with the tabular right-to-left parsing algorithm above
is that it computes many results that are never needed.
An additional inconvenience is that
we must carefully determine the order
in which the results for a particular column are computed,
so that parsing functions
such as {\tt pAdditive} and {\tt pMultitive}
that depend on other results from the same column
will work correctly.

{\em Packrat parsing} is essentially
a lazy version of the tabular algorithm
that solves both of these problems.
A packrat parser computes results only as they are needed,
in the same order
as the original recursive descent parser would.
However, once a result is computed for the first time,
it is stored for future use by subsequent calls.

A non-strict functional programming language such as Haskell
provides an ideal implementation platform for a packrat parser.
In fact, packrat parsing in Haskell is particularly efficient
because it does not require arrays or any other explicit lookup structures
other than the language's ordinary algebraic data types.

First we will need a new type
to represent a single column of the parsing result matrix,
which we will call {\tt Derivs} (``derivations'').
This type is merely a tuple
with one component for each nonterminal in the grammar.
Each component's type
is the result type of the corresponding parse function.
The {\tt Derivs} type also contains one additional component,
which we will call {\tt dvChar},
to represent ``raw'' characters of the input string
as if they were themselves the results of some parsing function.
The {\tt Derivs} type for our example parser
can be conveniently declared in Haskell as follows:

\begin{small}
\begin{verbatim}
data Derivs = Derivs {
                dvAdditive  :: Result Int,
                dvMultitive :: Result Int,
                dvPrimary   :: Result Int,
                dvDecimal   :: Result Int,
                dvChar      :: Result Char}
\end{verbatim}
\end{small}

This Haskell syntax declares the type {\tt Derivs}
to have a single constructor, also named {\tt Derivs},
with five components of the specified types.
The declaration also automatically creates
a corresponding data-accessor function for each component:
{\tt dvAdditive} can be used as a function
of type {\tt Derivs $\rightarrow$ Result Int},
which extracts the first component of a {\tt Derivs} tuple,
and so on.

Next we modify the {\tt Result} type
so that the ``remainder'' component of a success result
is not a plain {\tt String},
but is instead an instance of {\tt Derivs}:

\begin{small}
\begin{verbatim}
data Result v = Parsed v Derivs
              | NoParse
\end{verbatim}
\end{small}

The {\tt Derivs} and {\tt Result} types are now mutually recursive:
the success results in one {\tt Derivs} instance
act as links to other {\tt Derivs} instances.
These result values in fact provide the {\em only} linkage we need
between different columns in the matrix of parsing results.

\com{
\begin{small}
\begin{center}
\begin{verbatim}
pAdditive  d = ... (uses dvAdditive, pMultitive, dvChar) ...
pMultitive d = ... (uses dvMultitive, dvPrimary, dvChar) ...
pPrimary   d = ... (uses dvAdditive, dvDecimal, dvChar) ...
pDecimal   d = ... (uses dvChar) ...
\end{verbatim}
\end{center}
\end{small}
}

Now we modify the original recursive-descent parsing functions
so that each takes a {\tt Derivs}
instead of a {\tt String} as its parameter:

\begin{small}
\begin{verbatim}
pAdditive  :: Derivs -> Result Int
pMultitive :: Derivs -> Result Int
pPrimary   :: Derivs -> Result Int
pDecimal   :: Derivs -> Result Int
\end{verbatim}
\end{small}

Wherever one of the original parse functions
examined input characters directly,
the new parse function instead refers to the {\tt dvChar} component
of the {\tt Derivs} object.
Wherever one of the original functions
made a recursive call to itself or another parse function,
in order to match a nonterminal in the grammar,
the new parse function instead instead uses
the {\tt Derivs} accessor function corresponding to that nonterminal.
Sequences of terminals and nonterminals are matched
by following chains of success results
through multiple {\tt Derivs} instances.
For example, the new {\tt pAdditive} function
uses the {\tt dvMultitive}, {\tt dvChar}, and {\tt dvAdditive} accessors
as follows,
without making any direct recursive calls:

\begin{small}
\begin{verbatim}
-- Parse an additive-precedence expression
pAdditive :: Derivs -> Result Int
pAdditive d = alt1 where

    -- Additive <- Multitive '+' Additive
    alt1 = case dvMultitive d of
             Parsed vleft d' ->
               case dvChar d' of
                 Parsed '+' d'' ->
                   case dvAdditive d'' of
                     Parsed vright d''' ->
                       Parsed (vleft + vright) d'''
                     _ -> alt2
                 _ -> alt2
             _ -> alt2

    -- Additive <- Multitive
    alt2 = dvMultitive d
\end{verbatim}
\end{small}

Finally, we create a special ``top-level'' function, {\tt parse},
to produce instances of the {\tt Derivs} type
and ``tie up'' the recursion between all of the individual parsing functions:

\begin{small}
\begin{verbatim}
-- Create a result matrix for an input string
parse :: String -> Derivs
parse s = d where
    d    = Derivs add mult prim dec chr
    add  = pAdditive d
    mult = pMultitive d
    prim = pPrimary d
    dec  = pDecimal d
    chr  = case s of
             (c:s') -> Parsed c (parse s')
             [] -> NoParse
\end{verbatim}
\end{small}

The ``magic'' of the packrat parser is in this doubly-recursive function.
The first level of recursion is produced by
the {\tt parse} function's reference to itself
within the {\tt case} statement.
This relatively conventional form of recursion is used
to iterate over the input string one character at a time,
producing one {\tt Derivs} instance for each input position.
The final {\tt Derivs} instance, representing the empty string,
is assigned a {\tt dvChar} result of {\tt NoParse},
which effectively terminates the list of columns in the result matrix.

The second level of recursion is via the symbol {\tt d}.
This identifier names the {\tt Derivs} instance
to be constructed and returned by the {\tt parse} function,
but it is also the parameter
to each of the individual parsing functions.
These parsing functions, in turn,
produce the rest of the components
forming this very {\tt Derivs} object.

This form of {\em data recursion} of course
works only in a non-strict language,
which allow some components of an object to be accessed
before other parts of the same object are available.
For example, in any {\tt Derivs} instance created by the above function,
the {\tt dvChar} component can be accessed
before any of the other components of the tuple are available.
Attempting to access the {\tt dvDecimal} component of this tuple
will cause {\tt pDecimal} to be invoked,
which in turn uses the {\tt dvChar} component
but does not require any of the other ``higher-level'' components.
Accessing the {\tt dvPrimary} component
will similarly invoke {\tt pPrimary},
which may access {\tt dvChar} and {\tt dvAdditive}.
Although in the latter case {\tt pPrimary}
is accessing a ``higher-level'' component,
doing so does not create a cyclic dependency in this case
because it only ever invokes {\tt dvAdditive}
on a {\em different} {\tt Derivs} object from the one it was called with:
namely the one for the position following the opening parenthesis.
Every component of every {\tt Derivs} object produced by {\tt parse}
can be lazily evaluated in this fashion.

\begin{figure}
\centerline{\epsfig{file=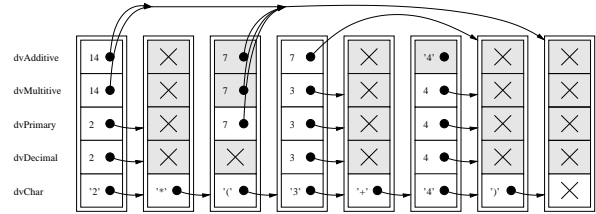, scale=0.35}}
\caption{Illustration of {\tt Derivs} data structure produced
	 by parsing the string `{\tt 2*(3+4)}'}
\label{derivs-diagram}
\end{figure}

Figure~\ref{derivs-diagram} illustrates the data structure
produced by the parser for the example input text `{\tt 2*(3+4)}',
as it would appear in memory under a modern functional evaluator
after fully reducing every cell.
Each vertical column represents a {\tt Derivs} instance
with its five {\tt Result} components.
For results of the form `{\tt Parsed $v$ $d$}',
the semantic value $v$ is shown in the appropriate cell,
along with an arrow representing the ``remainder'' pointer
leading to another {\tt Derivs} instance in the matrix.
In any modern lazy language implementation
that properly preserves sharing relationships during evaluation,
the arrows in the diagram will literally correspond to pointers in the heap,
and a given cell in the structure will never be evaluated twice.
Shaded boxes represent cells
that would never be evaluated at all
in the likely case that
the {\tt dvAdditive} result in the leftmost column
is the only value ultimately needed by the application.

This illustration should make it clear
why this algorithm can run in $O(n)$ time under a lazy evaluator
for an input string of length $n$.
The top-level {\tt parse} function is the {\em only} function
that creates instances of the {\tt Derivs} type,
and it always creates exactly $n+1$ instances.
The parse functions only access entries in this structure
instead of making direct calls to each other,
and each function examines at most a fixed number of other cells
while computing a given result.
Since the lazy evaluator ensures that each cell is evaluated at most once,
the critical memoization property is provided
and linear parse time is guaranteed,
even though the order in which these results are evaluated
is likely to be completely different from the
tabular, right-to-left, bottom-to-top algorithm presented earlier.

\section{Extending the Algorithm}
\label{extensions}

The previous section provided
the basic principles and tools required to create a packrat parser,
but building parsers for real applications involves many additional details,
some of which are affected by the packrat parsing paradigm.
In this section we will explore
some of the more important practical issues,
while incrementally building on the example packrat parser
developed above.
We first examine the annoying but straightforward problem of left recursion.
Next we address the issue of lexical analysis,
seamlessly integrating this task into the packrat parser.
Finally, we explore the use of monadic combinators
to express packrat parsers more concisely.

\subsection{Left Recursion}

One limitation packrat parsing shares with other top-down schemes
is that it does not directly support {\em left recursion}.
For example, suppose we wanted
to add a subtraction operator to the above example
and have addition and subtraction be properly left-associative.
A natural approach would be
to modify the grammar rules for Additive expressions as follows,
and to change the parser accordingly:

\begin{center}
\begin{tabular}{lcl}
Additive & $\leftarrow$ & Additive `{\tt +}' Multitive \\
	 & $|$		& Additive `{\tt -}' Multitive \\
	 & $|$		& Multitive \\
\end{tabular}
\end{center}

In a recursive descent parser for this grammar,
the {\tt pAdditive} function would recursively invoke itself
with the same input it was provided,
and therefore would get into an infinite recursion cycle.
In a packrat parser for this grammar,
{\tt pAdditive} would attempt to access the {\tt dvAdditive} component
of {\em its own} {\tt Derivs} tuple---%
the same component it is supposed to compute---%
and thus would create a circular data dependency.
In either case the parser fails,
although the packrat parser's failure mode
might be viewed as slightly ``friendlier''
since modern lazy evaluators
often detect circular data dependencies at run-time
but cannot detect infinite recursion.

\com{
\begin{figure}[t]
\begin{small}
\begin{center}
\begin{verbatim}
data Derivs = Derivs {
        dvAdditive       :: Result Int,
        dvAdditiveSuffix :: Result (Int -> Int),
        ...}

-- Additive <- Multitive AdditiveSuffix
pAdditive :: Derivs -> Result Int
pAdditive d = case dvMultitive d of
                Parsed vl d' ->
                  case dvAdditiveSuffix d' of
                    Parsed suf d'' ->
                      Parsed (suf vl) d''
                    _ -> NoParse
                _ -> NoParse

pAdditiveSuffix :: Derivs -> Result (Int -> Int)
pAdditiveSuffix d = alt1 where

  -- AdditiveSuffix <- '+' Multitive AdditiveSuffix
  alt1 = case dvChar d of
           Parsed '+' d' ->
             case dvMultitive d' of
               Parsed vr d'' ->
                 case dvAdditiveSuffix d'' of
                   Parsed suf d''' ->
                     Parsed (\vl -> suf (vl + vr))
                            d'''
                   _ -> alt2
               _ -> alt2
           _ -> alt2

  -- AdditiveSuffix <- '-' Multitive AdditiveSuffix
  alt2 = ...

  -- AdditiveSuffix <- <empty>
  alt3 = Parsed (\v -> v) d
\end{verbatim}
\end{center}
\end{small}
\caption{Parsing left-associative operators}
\label{triv-left}
\end{figure}
}

Fortunately,
a left-recursive grammar can always be rewritten
into an equivalent right-recursive one~\cite{aho86compilers},
and the desired left-associative semantic behavior is easily reconstructed
using higher-order functions as intermediate parser results.
For example,
to make Additive expressions left-associative in the example parser,
we can split this rule into two nonterminals, Additive and AdditiveSuffix.
The {\tt pAdditive} function
recognizes a single Multitive expression
followed by an AdditiveSuffix:

\begin{small}
\begin{verbatim}
pAdditive :: Derivs -> Result Int
pAdditive d = case dvMultitive d of
                Parsed vl d' ->
                  case dvAdditiveSuffix d' of
                    Parsed suf d'' ->
                      Parsed (suf vl) d''
                    _ -> NoParse
                _ -> NoParse
\end{verbatim}
\end{small}

The {\tt pAdditiveSuffix} function
collects infix operators and right-hand-side operands,
and builds a semantic value of type `Int $\rightarrow$ Int',
which takes a left-hand-side operand and produces a result:

\begin{small}
\begin{verbatim}
pAdditiveSuffix :: Derivs -> Result (Int -> Int)
pAdditiveSuffix d = alt1 where

  -- AdditiveSuffix <- '+' Multitive AdditiveSuffix
  alt1 = case dvChar d of
           Parsed '+' d' ->
             case dvMultitive d' of
               Parsed vr d'' ->
                 case dvAdditiveSuffix d'' of
                   Parsed suf d''' ->
                     Parsed (\vl -> suf (vl + vr))
                            d'''
                   _ -> alt2
               _ -> alt2
           _ -> alt2

  -- AdditiveSuffix <- <empty>
  alt3 = Parsed (\v -> v) d
\end{verbatim}
\end{small}


\subsection{Integrated Lexical Analysis}
\label{lexical}

Traditional parsing algorithms usually assume that
the ``raw'' input text has already been partially digested
by a separate {\em lexical analyzer} into a stream of tokens.
The parser then treats these tokens as atomic units
even though each may represent multiple consecutive input characters.
This separation is usually necessary
because conventional linear-time parsers 
can only use primitive terminals in their lookahead decisions
and cannot refer to higher-level nonterminals.
This limitation
was explained in Section~\ref{prediction}
for predictive top-down parsers,
but bottom-up LR parsers also depend on
a similar token-based lookahead mechanism
sharing the same problem.
If a parser can only use atomic tokens in its lookahead decisions,
then parsing becomes much easier if those tokens represent
whole keywords, identifiers, and literals
rather than raw characters.

Packrat parsing suffers from no such lookahead limitation, however.
Because a packrat parser reflects a true backtracking model,
decisions between alternatives in one parsing function
can depend on {\em complete results} produced by other parsing functions.
For this reason, lexical analysis can be integrated seamlessly
into a packrat parser with no special treatment.

To extend the packrat parser example
with ``real'' lexical analysis,
we add some new nonterminals to the {\tt Derivs} type:

\begin{small}
\begin{verbatim}
data Derivs = Derivs {
                -- Expressions
                dvAdditive   :: Result Int,
                ...

                -- Lexical tokens
                dvDigits     :: Result (Int, Int),
                dvDigit      :: Result Int,
                dvSymbol     :: Result Char,
                dvWhitespace :: Result (),

                -- Raw input
                dvChar       :: Result Char}
\end{verbatim}
\end{small}

The {\tt pWhitespace} parse function
consumes any whitespace that may separate lexical tokens:

\begin{small}
\begin{verbatim}
pWhitespace :: Derivs -> Result ()
pWhitespace d = case dvChar d of
                  Parsed c d' ->
                    if isSpace c
                    then pWhitespace d'
                    else Parsed () d
                  _ -> Parsed () d
\end{verbatim}
\end{small}

In a more complete language,
this function might have the task of eating comments as well.
Since the full power of packrat parsing is available for lexical analysis,
comments could have a complex hierarchical structure of their own,
such as nesting or markups for literate programming.
Since syntax recognition is not broken into a unidirectional pipeline,
lexical constructs can even refer ``upwards''
to higher-level syntactic elements.
For example, a language's syntax could allow
identifiers or code fragments embedded within comments
to be demarked so the parser can find and analyze them
as actual expressions or statements,
making intelligent software engineering tools more effective.
Similarly, escape sequences in string literals
could contain generic expressions
representing static or dynamic substitutions.

The {\tt pWhitespace} example also illustrates
how commonplace {\em longest-match} disambiguation rules
can be easily implemented in a packrat parser,
even though they are difficult to express in a pure context-free grammar.
More sophisticated decision and disambiguation strategies
are easy to implement as well,
including general {\em syntactic predicates}~\cite{parr94adding},
which influence parsing decisions based on syntactic lookahead information
without actually consuming input text.
For example, the useful {\em followed-by} and {\em not-followed-by} rules
allow a parsing alternative to be used
only if the text matched by that alternative
is (or is not) followed by text matching some other arbitrary nonterminal.
Syntactic predicates of this kind
require unlimited lookahead in general
and are therefore outside the capabilities
of most other linear-time parsing algorithms.

Continuing with the lexical analysis example,
the function {\tt pSymbol} recognizes ``operator tokens''
consisting of an operator character
followed by optional whitespace:

\begin{small}
\begin{verbatim}
-- Parse an operator followed by optional whitespace
pSymbol :: Derivs -> Result Char
pSymbol d = case dvChar d of
                Parsed c d' ->
                  if c `elem` "+-*/%()"
                  then case dvWhitespace d' of
                         Parsed _ d'' -> Parsed c d''
                         _ -> NoParse
                  else NoParse
                _ -> NoParse
\end{verbatim}
\end{small}

Now we modify the higher-level parse functions for expressions
to use {\tt dvSymbol} instead of {\tt dvChar}
to scan for operators and parentheses.
For example, {\tt pPrimary} can be implemented as follows:

\begin{small}
\begin{verbatim}
-- Parse a primary expression
pPrimary :: Derivs -> Result Int
pPrimary d = alt1 where

    -- Primary <- '(' Additive ')'
    alt1 = case dvSymbol d of
             Parsed '(' d' ->
               case dvAdditive d' of
                 Parsed v d'' ->
                   case dvSymbol d'' of
                     Parsed ')' d''' -> Parsed v d'''
                     _ -> alt2
                 _ -> alt2
             _ -> alt2

    -- Primary <- Decimal
    alt2 = dvDecimal d
\end{verbatim}
\end{small}

This function demonstrates how parsing decisions can depend
not only on the {\em existence} of a match at a given position
for a nonterminal such as Symbol,
but also on the {\em semantic value} associated with that nonterminal.
In this case, even though all symbol tokens
are parsed together and treated uniformly by {\tt pSymbol},
other rules such as {\tt pPrimary}
can still distinguish between particular symbols.
In a more sophisticated language with multi-character operators,
identifiers, and reserved words,
the semantic values produced by the token parsers
might be of type {\tt String} instead of {\tt Char},
but these values can be matched in the same way.
Such dependencies of syntax on semantic values,
known as {\em semantic predicates}~\cite{parr94adding},
provide an extremely powerful and useful capability in practice.
As with syntactic predicates,
semantic predicates require unlimited lookahead in general
and cannot be implemented by conventional parsing algorithms
without giving up their linear time guarantee.


\subsection{Monadic Packrat Parsing}
\label{monadic}

A popular method of constructing parsers
in functional languages such as Haskell
is using monadic combinators~\cite{hutton98monadic, leijen01parsec}.
Unfortunately, the monadic approach usually comes with a performance penalty,
and with packrat parsing this tradeoff presents a difficult choice.
Implementing a packrat parser as described so far
assumes that the set of nonterminals and their corresponding result types
is known statically,
so that they can be bound together in a single fixed tuple
to form the {\tt Derivs} type.
Constructing entire packrat parsers dynamically from other packrat parsers
via combinators
would require making the {\tt Derivs} type a dynamic lookup structure,
associating a variable set of nonterminals with corresponding results.
This approach would be much slower and less space-efficient.

A more practical strategy,
which provides most of the convenience of combinators
with a less significant performance penalty,
is to use monads to define
the individual parsing {\em functions} comprising a packrat parser,
while keeping the {\tt Derivs} type and the ``top-level'' recursion
statically implemented as described earlier.

Since we would like our combinators
to build the parse functions we need directly,
the obvious method would be to make the combinators
work with a simple type alias:

\begin{small}
\begin{verbatim}
type Parser v = Derivs -> Result v
\end{verbatim}
\end{small}

Unfortunately,
in order to take advantage of Haskell's useful {\tt do} syntax,
the combinators must use a type of the special class {\tt Monad},
and simple aliases cannot be assigned type classes.
We must instead wrap the parsing functions
with a ``real'' user-defined type:

\begin{small}
\begin{verbatim}
newtype Parser v = Parser (Derivs -> Result v)
\end{verbatim}
\end{small}

We can now implement Haskell's standard sequencing ({\tt >>=}),
result-producing ({\tt return}),
and error-producing combinators:

\begin{small}
\begin{verbatim}
instance Monad Parser where

        (Parser p1) >>= f2 = Parser pre
                where pre d = post (p1 d)
                      post (Parsed v d') = p2 d'
                        where Parser p2 = f2 v
                      post (NoParse) = NoParse

        return x = Parser (\d -> Parsed x d)

        fail msg = Parser (\d -> NoParse)
\end{verbatim}
\end{small}

Finally, for parsing we need an alternation combinator:

\begin{small}
\begin{verbatim}
(<|>) :: Parser v -> Parser v -> Parser v
(Parser p1) <|> (Parser p2) = Parser pre
                where pre d = post d (p1 d)
                      post d NoParse = p2 d
                      post d r = r
\end{verbatim}
\end{small}

With these combinators
in addition to a trivial one to recognize specific characters,
the {\tt pAdditive} function in the original packrat parser example
can be written as follows:

\begin{small}
\begin{verbatim}
Parser pAdditive =
            (do vleft <- Parser dvMultitive
                char '+'
                vright <- Parser dvAdditive
                return (vleft + vright))
        <|> (do Parser dvMultitive)
\end{verbatim}
\end{small}

It is tempting to build additional combinators
for higher-level idioms such as repetition and infix expressions.
However, using iterative combinators within packrat parsing functions
violates the assumption that each cell in the result matrix
can be computed in constant time
once the results from any other cells it depends on are available.
Iterative combinators effectively create ``hidden'' recursion
whose intermediate results are not memoized in the result matrix,
potentially making the parser run in super-linear time.
This problem is not necessarily serious in practice,
as the results in Section~\ref{results} will show,
but it should be taken into account
when using iterative combinators.

The on-line examples for this paper
include a full-featured monadic combinator library
that can be used to build large packrat parsers conveniently.
This library is substantially inspired by {\sc Parsec}~\cite{leijen01parsec},
though the packrat parsing combinators are much simpler
since they do not have to implement lexical analysis as a separate phase
or implement the one-token-lookahead prediction mechanism
used by traditional top-down parsers.
The full combinator library
provides a variety of ``safe'' constant-time combinators,
as well as a few ``dangerous'' iterative ones,
which are convenient but not necessary to construct parsers.
The combinator library can be used simultaneously
by multiple parsers with different {\tt Derivs} types,
and supports user-friendly error detection and reporting.


\section{Comparison with LL and LR Parsing}
\label{comparison}

Whereas the previous sections have served as a tutorial
on {\em how} to construct a packrat parser,
for the remaining sections we turn to the issue
of {\em when} packrat parsing is useful in practice.
This section informally explores the language recognition power
of packrat parsing in more depth,
and clarifies its relationship to traditional linear-time algorithms
such as LL($k$) and LR($k$).

Although LR parsing is commonly seen as ``more powerful''
than limited-lookahead top-down or LL parsing,
the class of languages these parsers can recognize
is the same~\cite{aho72parsing}.
As Pepper points out~\cite{pepper99transform},
LR parsing can be viewed simply as LL parsing
with the grammar rewritten so as to eliminate left recursion
and to delay all important parsing decisions as long as possible.
The result is that LR provides more flexibility
in the way grammars can be expressed,
but no actual additional recognition power.
For this reason, we will treat LL and LR parsers here
as being essentially equivalent.
\com{
In particular,
most practical LR parsers use the same basic lookahead scheme as LL parsers do,
allowing decisions to take into account
only a single primitive token (terminal)
beyond the current position of the parser.
Since these transformations are fairly involved,
in practice they are automated by a grammar compiler
such as the classic YACC.
A practical ``packrat parsing compiler compiler''
could easily apply similar transformations to TDPL grammars
to achieve similar benefits,
most notably the ability to express left recursion directly,
though this has not yet been done.
}

\subsection{Lookahead}

The most critical practical difference
between packrat parsing and LL/LR parsing is the lookahead mechanism.
A packrat parser's decisions at any point
can be based on all the text up to the end of the input string.
Although the computation of an individual result in the parsing matrix
can only perform a constant number of ``basic operations,''
these basic operations
include following forward pointers in the parsing matrix,
each of which can skip over a large amount of text at once.
Therefore,
while LL and LR parsers can only look ahead
a constant number of {\em terminals} in the input,
packrat parsers can look ahead
a constant number of {\em terminals and nonterminals} in any combination.
This ability for parsing decisions to take arbitrary nonterminals into account
is what gives packrat parsing its unlimited lookahead capability.

\com{
From a viewpoint of computational complexity,
packrat parsing inherently requires, and takes advantage of,
a more powerful computational model than LL or LR parsing.
Whereas LL and LR parsers can be implemented
by a simple stack machine
or {\em deterministic push-down automata}~\cite{aho86compilers},
packrat parsing inherently requires a machine with random-access memory.
}

\com{
Furthermore,
a packrat parser implemented in a general language such as Haskell,
as opposed to a restricted formal language such as TDPL,
is not limited to using only the {\em existence} of derivations
for particular terminals or nonterminals farther right in the input
when computing a particular derivation or deciding between alternatives;
it can also use the {\em semantic values} associated with those derivations
without restriction.
This flexibility allows the parser to handle
many kinds of ``infinite'' grammars:
i.e., grammars that effectively have an infinite number of symbols and rules.
A simple contrived example is a parser for the language
$\{$`{\tt 1a}', `{\tt 2aa}', `{\tt 3aaa}', $\dots\}$.
Predictive parsers can also handle
some kinds of infinite grammars as well in this way,
but with much more severe restrictions
due to the fixed-lookahead constraint.
}

To illustrate the difference in language recognition power,
the following grammar is not LR($k$) for any $k$,
but is not a problem for a packrat parser:

\begin{center}
\begin{tabular}{ccl}
S & $\leftarrow$ & A $|$ B \\
A & $\leftarrow$ & x A y $|$ x z y \\
B & $\leftarrow$ & x B y y $|$ x z y y \\
\end{tabular}
\end{center}

Once an LR parser has encountered the `{\tt z}'
and the first following `{\tt y}'
in a string in the above language,
it must decide immediately whether to start reducing via nonterminal A or B,
but there is no way for it to make this decision
until as many `{\tt y}'s have been encountered
as there were `{\tt x}'s on the left-hand side.
A packrat parser, on the other hand,
essentially operates in a speculative fashion,
producing derivations for nonterminals A and B {\em in parallel}
while scanning the input.
The ultimate decision between A and B is effectively delayed
until the {\em entire} input string has been parsed,
where the decision is merely a matter of checking
which nonterminal has a success result at that position.
Mirroring the above grammar left to right does not change the situation,
making it clear that the difference
is not merely some side-effect of the fact that
LR scans the input left-to-right
whereas packrat parsing seems to operate in reverse.

\subsection{Grammar Composition}

The limitations of LR parsing due to fixed lookahead
are frequently felt when designing parsers for practical languages,
and many of these limitations stem from the fact that
LL and LR grammars are not cleanly {\em composable}.
\com{
For example, as mentioned before,
The primary reason it has traditionally been impractical
to integrate lexical and hierarchical analysis in a single phase
is because LR grammars are not {\em composable}.
If two LR grammars
(e.g., a lexical grammar and an expression grammar)
are combined so that nonterminals in one
are substituted for the terminals in the other,
the result is usually {\em not} an LR grammar.
}
For example, the following grammar
represents a simple language with expressions and assignment,
which only allows simple identifiers on the left side of an assignment:

\begin{center}
\begin{tabular}{ccl}
S & $\leftarrow$ & R $|$ ID `{\tt =}' R \\
R & $\leftarrow$ & A $|$ A EQ A $|$ A NE A \\
A & $\leftarrow$ & P $|$ P `{\tt +}' P $|$ P `{\tt -}' P \\
P & $\leftarrow$ & ID $|$ `{\tt (}' R `{\tt )}' \\
\end{tabular}
\end{center}

If the symbols ID, EQ, and NE are terminals---%
i.e., atomic tokens produced by a separate lexical analysis phase---%
then an LR(1) parser has no trouble with this grammar.
However, if we try to integrate this tokenization into the parser itself
with the following simple rules,
the grammar is no longer LR(1):

\begin{center}
\begin{tabular}{ccl}
ID & $\leftarrow$ & '{\tt a}' $|$ '{\tt a}' ID \\
EQ & $\leftarrow$ & '{\tt =}' '{\tt =}' \\
NE & $\leftarrow$ & '{\tt !}' '{\tt =}' \\
\end{tabular}
\end{center}

The problem is that after scanning an identifier,
an LR parser must decide immediately
whether it is a primary expression or the left-hand side of an assignment,
based only on the immediately following token.
But if this token is an `{\tt =}',
the parser has no way of knowing whether it is an assignment operator
or the first half of an `{\tt ==}' operator.
In this particular case the grammar could be parsed by an LR(2) parser.
In practice LR($k$) and even LALR($k$) parsers
are uncommon for $k > 1$.
Recently developed extensions
to the traditional left-to-right parsing algorithms
improve the situation somewhat~%
\cite{salomon89scannerless,parr93thesis,parr95antlr},
but they still cannot provide unrestricted lookahead capability
while maintaining the linear time guarantee.

Even when lexical analysis is separated from parsing,
the limitations of LR parsers often surface in other practical situations,
frequently as a result of seemingly innocuous changes to an evolving grammar.
For example, suppose we want to add simple array indexing to the language above,
so that array indexing operators can appear
on either the left or right side of an assignment.
One possible approach is to add a new nonterminal, L,
to represent left-side or ``lvalue'' expressions,
and incorporate the array indexing operator into both types of expressions
as shown below:

\begin{center}
\begin{tabular}{ccl}
S & $\leftarrow$ & R $|$ L `{\tt =}' R \\
R & $\leftarrow$ & A $|$ A EQ A $|$ A NE A \\
A & $\leftarrow$ & P $|$ P `{\tt +}' P $|$ P `{\tt -}' P \\
P & $\leftarrow$ & ID $|$ `{\tt (}' R `{\tt )}' $|$ P `{\tt [}' A `{\tt ]}' \\
L & $\leftarrow$ & ID $|$ `{\tt (}' L `{\tt )}' $|$ L `{\tt [}' A `{\tt ]}' \\
\end{tabular}
\end{center}

Even if the ID, EQ, and NE symbols are again treated as terminals,
this grammar is not LR($k$) for any $k$,
because after the parser sees an identifier
it must immediately decide whether it is part of a P or L expression,
but it has no way of knowing this
until any following array indexing operators have been fully parsed.
Again, a packrat parser has no trouble with this grammar
because it effectively evaluates the P and L alternatives ``in parallel''
and has complete derivations to work with
(or the knowledge of their absence)
by the time the critical decision needs to be made.

In general, grammars for packrat parsers are composable
because the lookahead a packrat parser uses
to make decisions between alternatives
can take account of arbitrary nonterminals,
such as EQ in the first example or P and L in the second.
Because a packrat parser
does not give ``primitive'' syntactic constructs (terminals)
any special significance as an LL or LR parser does,
any terminal or fixed sequence of terminals appearing in a grammar
can be substituted with a nonterminal
without ``breaking'' the parser.
This substitution capability
gives packrat parsing greater composition flexibility.


\subsection{Recognition Limitations}

Given that a packrat parser can recognize
a broader class of languages in linear time
than either LL($k$) or LR($k$) algorithms,
what kinds of grammars {\em can't} a packrat parser recognize?
Though the precise theoretical capabilities of the algorithm
have not been thoroughly characterized,
the following trivial and unambiguous context-free grammar
provides an example
that proves just as troublesome for a packrat parser
as for an LL or LR parser:

\begin{center}
\begin{tabular}{ccl}
S & $\leftarrow$ & x S x $|$ x \\
\end{tabular}
\end{center}

The problem with this grammar for both kinds of parsers is that,
while scanning a string of `{\tt x}'s---%
left-to-right in the LR case or right-to-left in the packrat case---%
the algorithm would somehow have to ``know'' in advance
where the middle of the string is
so that it can apply the second alternative at that position
and then ``build outwards'' using the first alternative
for the rest of the input stream.
But since the stream is completely homogeneous,
there is no way for the parser to find the middle
until the entire input has been parsed.
This grammar therefore provides an example, albeit contrived,
requiring a more general, non-linear-time
CFG parsing algorithm.

\com{
Of course, in this case it is easy to write an LL(1) grammar
(or even a regular expression)
that recognizes the same class of languages---%
namely the set of odd-length strings of `{\tt x}'s---%
but the resulting parser would effectively derive less information
from the input stream than the original grammar does:
i.e., it would no longer ``find the middle.''
Furthermore,
the following context-free grammar probably cannot even be rewritten
into a TDPL grammar that recognizes the same class of languages:

\begin{center}
\begin{tabular}{ccl}
S & $\leftarrow$ & x S x $|$ A B \\
A & $\leftarrow$ & x A y $|$ x y \\
B & $\leftarrow$ & y B x $|$ y x \\
\end{tabular}
\end{center}

Intuitively, this grammar accepts strings
consisting of $(i+j)$ `{\tt x}'s,
followed by $(j+k)$ `{\tt y}'s,
followed by $(k+i)$ more `{\tt x}'s,
where $i,j,k \ge 1$.
Such a language could be recognized by a generalized algorithm
such as Earley's or Tomita's~\cite{tomita85efficient},
though not in linear time.
Of course this language could also trivially be recognized in linear time
by a general packrat parser written in Haskell
in which parsing decisions can be made
based on semantic values computed during the parsing process:
i.e., just count the number of characters in each sequence
and make sure the numbers add up at the end.
}

\section{Practical Issues and Limitations}
\label{issues}

Although packrat parsing is powerful and efficient enough
for many applications,
there are three main issues that can make it inappropriate in some situations.
First, packrat parsing is useful only
to construct {\em deterministic} parsers:
parsers that can produce at most one result.
Second, a packrat parser depends for its efficiency
on being mostly or completely {\em stateless}.
Finally, due to its reliance on memoization,
packrat parsing is inherently space-intensive.
These three issues are discussed in this section.

\subsection{Deterministic Parsing}

An important assumption we have made so far
is that each of the mutually recursive parsing functions
from which a packrat parser is built
will deterministically return {\em at most one result}.
If there are any ambiguities in the grammar the parser is built from,
then the parsing functions must be able to resolve them locally.
In the example parsers developed in this paper,
multiple alternatives have always been implicitly disambiguated
by the order in which they are tested:
the first alternative to match successfully is the one used,
independent of whether any other alternatives may also match.
This behavior is both easy to implement
and useful for performing longest-match
and other forms of explicit local disambiguation.
A parsing function could even
try all of the possible alternatives
and produce a failure result if more than one alternative matches.
What parsing functions in a packrat parser {\em cannot} do
is return {\em multiple} results
to be used in parallel or disambiguated later by some global strategy.

In languages designed for machine consumption,
the requirement that multiple matching alternatives be disambiguated locally
is not much of a problem in practice
because ambiguity is usually undesirable in the first place,
and localized disambiguation rules are preferred over global ones
because they are easier for humans to understand.
However, for parsing natural languages
or other grammars in which global ambiguity is expected,
packrat parsing is less likely to be useful.
Although a classic nondeterministic top-down parser
in which the parse functions return lists of results%
~\cite{wadler85how, fokker95functional}
could be memoized in a similar way,
the resulting parser would not be linear time,
and would likely be comparable to
existing tabular algorithms for ambiguous context-free grammars%
~\cite{aho72parsing, tomita85efficient}.
Since nondeterministic parsing is equivalent in computational complexity
to boolean matrix multiplication~\cite{lee02bmm},
a linear-time solution to this more general problem
is unlikely to be found.

\subsection{Stateless Parsing}
\label{stateless}

A second limitation of packrat parsing
is that it is fundamentally geared toward {\em stateless} parsing.
A packrat parser's memoization system assumes that
the parsing function for each nonterminal depends only on the input string,
and not on any other information accumulated during the parsing process.

Although pure context-free grammars are by definition stateless,
many practical languages require a notion of state while parsing
and thus are not really context-free.
For example, C and C++ require the parser
to build a table of type names incrementally as types are declared,
because the parser must be able
to distinguish type names from other identifiers
in order to parse subsequent text correctly.

Traditional top-down (LL) and bottom-up (LR) parsers
have little trouble maintaining state while parsing.
Since they perform only a single left-to-right scan of the input
and never look ahead more than one or at most a few tokens,
nothing is ``lost'' when a state change occurs.
A packrat parser, in contrast,
depends on statelessness
for the efficiency of its unlimited lookahead capability.
Although a stateful packrat parser can be constructed,
the parser must start building a new result matrix
each time the parsing state changes.
For this reason,
stateful packrat parsing may be impractical
if state changes occur frequently.
For more details on packrat parsing with state,
please refer to my master's thesis~\cite{ford02packrat}.

\subsection{Space Consumption}
\label{space}

Probably the most striking characteristic of a packrat parser
is the fact that it literally squirrels away
{\em everything} it has ever computed about the input text,
including the entire input text itself.
For this reason packrat parsing always has storage requirements equal to
some possibly substantial constant multiple of the input size.
In contrast,
LL($k$), LR($k$), and simple backtracking parsers can be designed
so that space consumption grows only with the {\em maximum nesting depth}
of the syntactic constructs appearing in the input,
which in practice is often orders of magnitude smaller
than the total size of the text.
Although LL($k$) and LR($k$) parsers for any non-regular language
still have linear space requirements in the worst case,
this ``average-case'' difference can be important in practice.

One way to reduce the space requirements of the derivations structure,
especially in parsers for grammars with many nonterminals,
is by splitting up the {\tt Derivs} type into multiple levels.
For example, suppose the nonterminals of a language
can be grouped into several broad categories,
such as lexical tokens, expressions, statements, and declarations.
Then the {\tt Derivs} tuple itself
might have only four components in addition to {\tt dvChar},
one for each of these nonterminal categories.
Each of these components is in turn a tuple
containing the results for all of the nonterminals in that category.
For the majority of the {\tt Derivs} instances,
representing character positions ``between tokens,''
none of the components representing the categories of nonterminals
will ever be evaluated,
and so only the small top-level object
and the unevaluated closures for its components occupy space.
Even for {\tt Derivs} instances corresponding to the beginning of a token,
often the results from only one or two categories will be needed
depending on what kind of language construct is located at that position.


Even with such optimizations
a packrat parser can consume
many times more working storage than the size of the original input text.
For this reason there are some application areas
in which packrat parsing is probably not the best choice.
For example, for parsing XML streams,
which have a fairly simple structure
but often encode large amounts
of relatively flat, machine-generated data,
the power and flexibility of packrat parsing is not needed
and its storage cost would not be justified.

On the other hand,
for parsing complex modern programming languages
in which the source code is usually written by humans
and the top priority is the power and expressiveness of the language,
the space cost of packrat parsing is probably reasonable.
Standard programming practice involves breaking up large programs
into modules of manageable size that can be independently compiled,
and the main memory sizes of modern machines
leave at least three orders of magnitude in ``headroom''
for expansion of a typical 10--100KB source file during parsing.
Even when parsing larger source files,
the working set may still be relatively small
due to the strong structural locality properties of realistic languages.
Finally,
since the entire derivations structure can be thrown away
after parsing is complete,
the parser's space consumption is likely to be irrelevant
if its result is fed into some other complex computation,
such as a global optimizer,
that requires as much space as the packrat parser used.
Section~\ref{results} will present evidence
that this space consumption can be reasonable in practice.

\section{Performance Results}
\label{results}

Although a detailed empirical analysis of packrat parsing
is outside the scope of this paper,
it is helpful to have some idea of how a packrat parser
is likely to behave in practice
before committing to a new and unfamiliar parsing paradigm.
For this reason,
this section presents a few experimental results
with realistic packrat parsers running on real source files.
For more detailed results, 
please refer to my master's thesis~\cite{ford02packrat}.

\subsection{Space Efficiency}

The first set of tests measure the space efficiency
of a packrat parser for the Java\footnote{
	Java is a trademark of Sun Microsystems, Inc.}
programming language.
I chose Java for this experiment because it has a rich and complex grammar,
but nevertheless adopts a fairly clean syntactic paradigm,
not requiring the parser to keep state about declared types
as C and C++ parsers do,
or to perform special processing between lexical and hierarchical analysis
as Haskell's layout scheme requires.

The experiment uses two different versions of this Java parser.
Apart from a trivial preprocessing stage
to canonicalize line breaks and Java's Unicode escape sequences,
lexical analysis for both parsers is fully integrated
as described in Section~\ref{lexical}.
One parser uses monadic combinators in its lexical analysis functions,
while the other parser relies only on primitive pattern matching.
Both parsers use monadic combinators
to construct all higher-level parsing functions.
Both parsers also use the technique described in Section~\ref{space}
of splitting the {\tt Derivs} tuple into two levels,
in order to increase modularity and reduce space consumption.
The parsers were compiled with the Glasgow Haskell Compiler\footnote{
	{\tt http://www.haskell.org/ghc/}}
version 5.04, with optimization and profiling enabled.
GHC's heap profiling system was used to measure live heap utilization,
which excludes unused heap space and collectible garbage
when samples are taken.

The test suite consists of 60 unmodified Java source files
from the Cryptix library\footnote{
	{\tt http://www.cryptix.org/}},
chosen because it includes
a substantial number of relatively large Java source files.
(Java source files are small on average
because the compilation model encourages programmers
to place each class definition in a separate file.)

\begin{figure}
\centerline{\epsfig{file=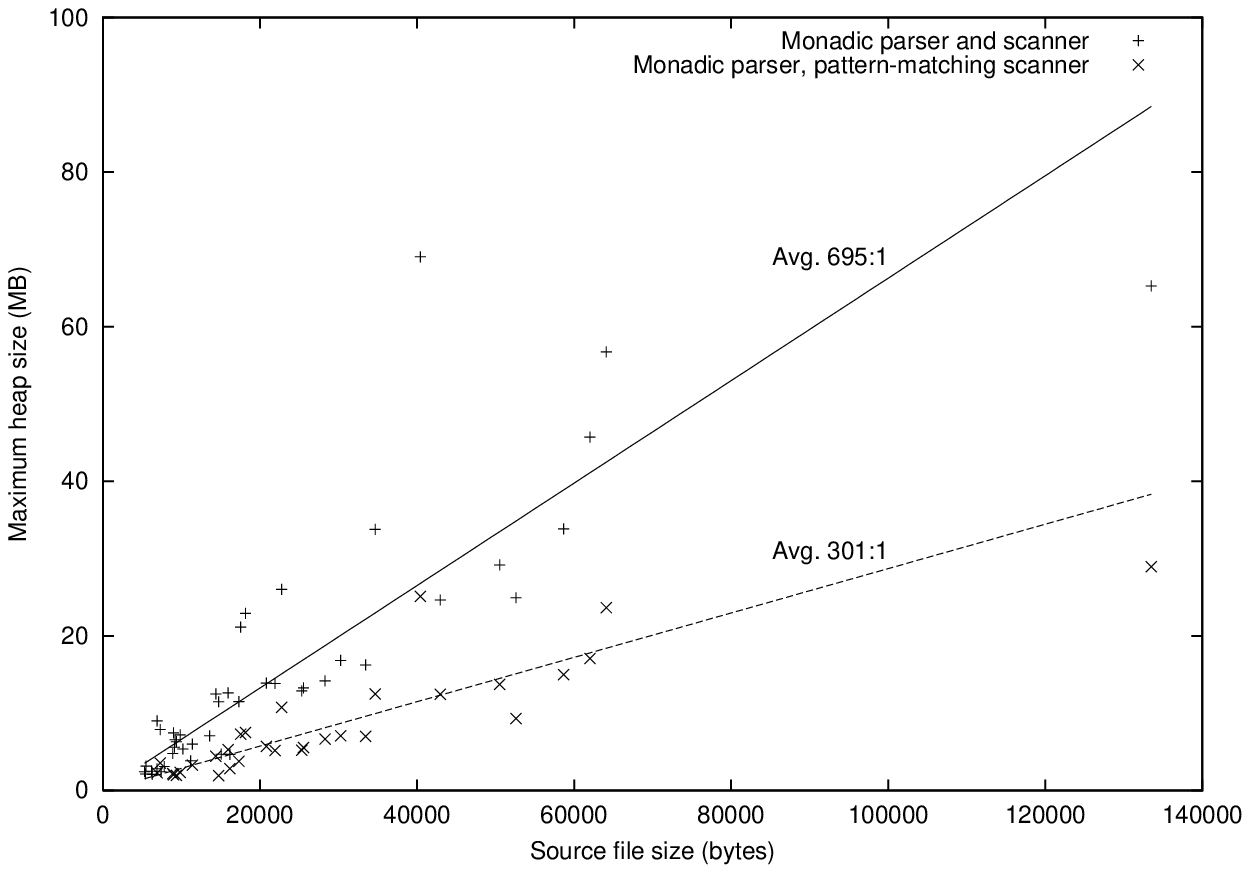, scale=0.6}}
\caption{Maximum heap size versus input size}
\label{heapsize}
\end{figure}

Figure~\ref{heapsize} shows a plot of
each parser's maximum live heap size
against the size of the input files being parsed.
Because some of the smaller source files were parsed so quickly
that garbage collection never occurred
and the heap profiling mechanism did not yield any samples,
the plot includes only 45 data points for the fully monadic parser,
and 31 data points for the hybrid parser
using direct pattern matching for lexical analysis.
Averaged across the test suite,
the fully monadic parser uses 695 bytes of live heap per byte of input,
while the hybrid parser
uses only 301 bytes of heap per input byte.
These results are encouraging:
although packrat parsing can consume a substantial amount of space,
a typical modern machine with 128MB or more of RAM
should have no trouble parsing source files up to 100-200KB.
Furthermore,
even though both parsers use some iterative monadic combinators,
which can break the linear time and space guarantee in theory,
the space consumption of the parsers
nevertheless appears to grow fairly linearly.

The use of monadic combinators clearly has a substantial penalty
in terms of space efficiency.
Modifying the parser to use direct pattern matching alone
may yield further improvement,
though the degree is difficult to predict
since the cost of lexical analysis often dominates the rest of the parser.
The lexical analysis portion of the hybrid parser
is about twice as long as the equivalent portion of the monadic parser,
suggesting that writing packrat parsers with pattern matching alone
is somewhat more cumbersome
but not unreasonable when efficiency is important.

\subsection{Parsing Performance}

\begin{figure}
\centerline{\epsfig{file=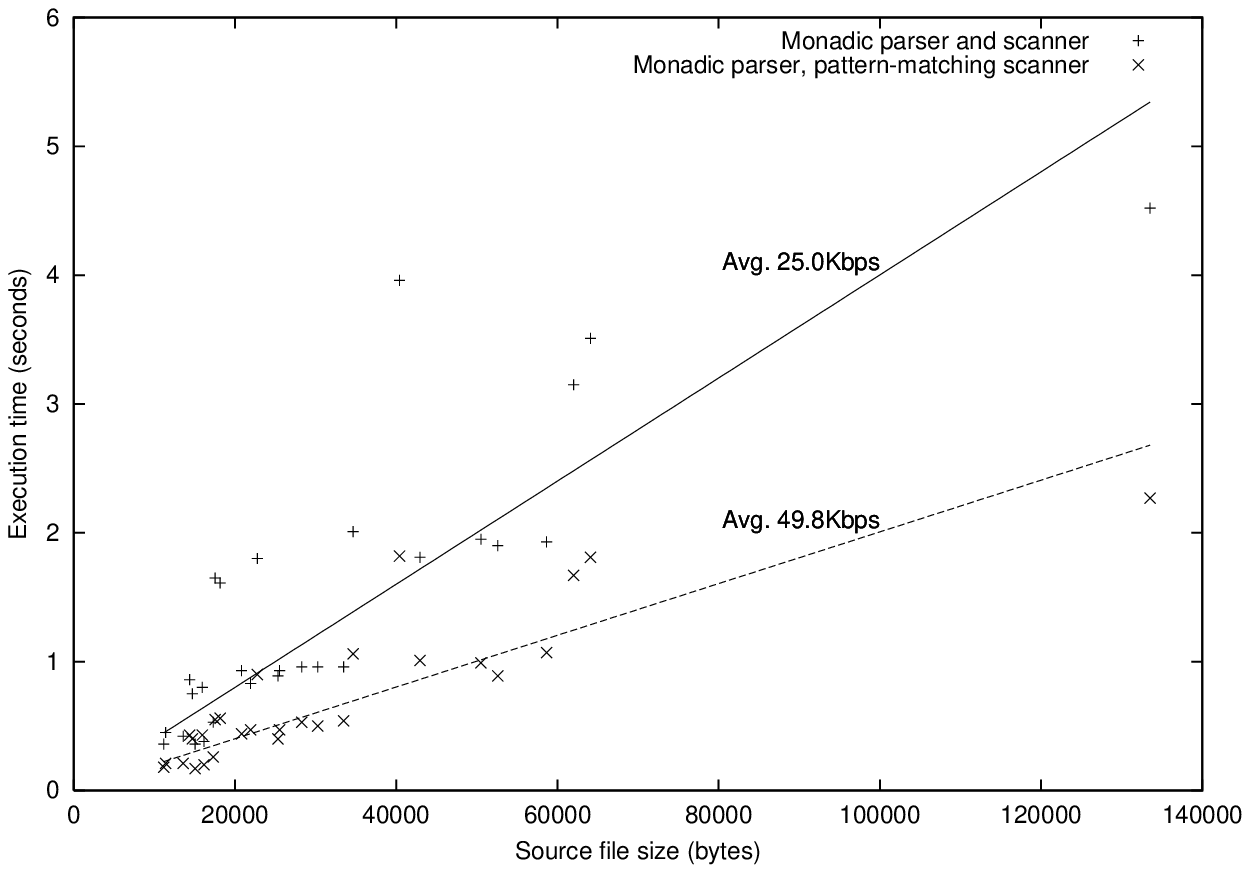, scale=0.6}}
\caption{Execution time versus input size}
\label{parsetime}
\end{figure}

The second experiment measures the absolute execution time
of the two packrat parsers.
For this test
the parsers were compiled by GHC 5.04 with optimization but without profiling,
and timed on a 1.28GHz AMD Athlon processor running Linux 2.4.17.
For this test I only used the 28 source files in the test suite
that were larger than 10KB,
because the smaller files were parsed so quickly
that the Linux {\tt time} command did not yield adequate precision.
Figure~\ref{parsetime} shows the resulting execution time
plotted against source file size.
On these inputs the fully monadic parser averaged 25.0 Kbytes per second
with a standard deviation of 8.6~KB/s,
while the hybrid parser averaged 49.8~KB/s
with a standard deviation of 16~KB/s.

In order to provide a legitimate performance comparison 
between packrat parsing and more traditional linear-time algorithms,
I converted a freely available YACC grammar for Java~\cite{bronnikov98java}
into a grammar for Happy\footnote{
	{\tt http://www.haskell.org/happy}},
an LR parser generator for Haskell.
Unfortunately,
GHC was unable to compile the 230KB Haskell source file
resulting from this grammar,
even without optimization and on a machine with 1GB of RAM.
(This difficulty incidentally lends credibility
to the earlier suggestion
that, in modern compilers, the temporary storage cost of a packrat parser
is likely to be exceeded by the storage cost of subsequent stages.)
Nevertheless,
the generated LR parser worked
under the Haskell interpreter Hugs.\footnote{
	{\tt http://www.haskell.org/hugs}}
Therefore, to provide a rough performance comparison,
I ran five of the larger Java sources through the LR and packrat parsers
under Hugs using an 80MB heap.
For fairness, I only compared the LR parser
against the slower, fully monadic packrat parser,
because the LR parser uses a monadic lexical analyzer
derived from the latter packrat parser.
The lexical analysis performance should therefore be comparable
and only the parsing algorithm is of primary importance.

Under Hugs, the LR parser consistently performs
approximately twice the number of reductions
and allocates 55\% more total heap storage.
(I could not find a way to profile {\em live} heap utilization under Hugs
instead of total allocation.)
The difference in real execution time varied widely however:
the LR parser took almost twice as long on smaller files
but performed about the same on the largest ones.
One probable reason for this variance
is the effects of garbage collection.
Since a running packrat parser will naturally have
a much higher ratio of live data to garbage than an LR parser over time,
and garbage collection both increases in overhead cost
and decreases in effectiveness (i.e., frees less space)
when there is more live data,
garbage collection is likely to penalize a packrat parser
more than an LR parser as the size of the source file increases.
Still, it is encouraging
that the packrat parser was able to outperform the LR parser
on all but the largest Java source files.

\section{Related Work}
\label{related}

This section briefly relates packrat parsing to relevant prior work.
For a more detailed analysis of packrat parsing
in comparison with other algorithms
please refer to my master's thesis~\cite{ford02packrat}.

Birman and Ullman~\cite{birman73parsing} first developed
the formal properties of deterministic parsing algorithms with backtracking.
This work was refined by Aho and Ullman~\cite{aho72parsing}
and classified as ``top-down limited backtrack parsing,''
in reference to the restriction
that each parsing function can produce at most one result
and hence backtracking is localized.
They showed this kind of parser,
formally known as a Generalized Top-Down Parsing Language (GTDPL) parser,
to be quite powerful.
A GTDPL parser can simulate any push-down automaton
and thus recognize any LL or LR language,
and it can even recognize some languages that are not context free.
Nevertheless, all ``failures'' such as those caused by left recursion
can be detected and eliminated from a GTDPL grammar,
ensuring that the algorithm is well-behaved.
Birman and Ullman also pointed out the possibility
of constructing linear-time GTDPL parsers through tabulation of results,
but this linear-time algorithm was apparently never put into practice,
no doubt because main memories were much more limited at the time
and compilers had to operate as streaming ``filters''
that could run in near-constant space.

Adams~\cite{adams91modular} recently resurrected GTDPL parsing
as a component of a modular language prototyping framework,
after recognizing its superior composability
in comparison with LR algorithms.
In addition, many practical top-down parsing libraries and toolkits,
including the popular ANTLR~\cite{parr95antlr}
and the {\sc Parsec} combinator library for Haskell~\cite{leijen01parsec},
provide similar limited backtracking capabilities
which the parser designer can invoke selectively
in order to overcome the limitations of predictive parsing.
However, all of these parsers implement backtracking
in the traditional recursive-descent fashion without memoization,
creating the danger of exponential worst-case parse time,
and thereby making it impractical
to rely on backtracking as a substitute for prediction
or to integrate lexical analysis with parsing.

The only prior known linear-time parsing algorithm
that effectively supports integrated lexical analysis,
or ``scannerless parsing,''
is the NSLR(1) algorithm originally created by Tai~\cite{tai79noncanonical}
and put into practice for this purpose
by Salomon and Cormack~\cite{salomon89scannerless}.
This algorithm extends the traditional LR class of algorithms
by adding limited support for making lookahead decisions based on nonterminals.
The relative power of packrat parsing with respect to NSLR(1) is unclear:
packrat parsing is less restrictive of rightward lookahead,
but NSLR(1) can also take leftward context into account.
In practice, NSLR(1) is probably more space-efficient,
but packrat parsing is simpler and cleaner.
Other recent scannerless parsers%
~\cite{visser97scannerless, brand02disambiguation}
forsake linear-time deterministic algorithms
in favor of more general but slower ambiguity-tolerant CFG parsing.

\section{Future Work}
\label{future}

While the results presented here demonstrate
the power and practicality of packrat parsing,
more experimentation is needed
to evaluate its flexibility, performance, and space consumption
on a wider variety of languages.
For example,
languages that rely extensively on parser state, such as C and C++,
as well as layout-sensitive languages such as ML and Haskell,
may prove more difficult for a packrat parser to handle efficiently.

On the other hand,
the syntax of a practical language is usually designed
with a particular parsing technology in mind.
For this reason,
an equally compelling question
is what new syntax design possibilities
are created by the ``free'' unlimited lookahead
and unrestricted grammar composition capabilities of packrat parsing.
Section~\ref{lexical} suggested a few simple extensions
that depend on integrated lexical analysis,
but packrat parsing may be even more useful
in languages with extensible syntax~\cite{cardelli94extensible}
where grammar composition flexibility is important.

Although packrat parsing is simple enough
to implement by hand in a lazy functional language,
there would still be practical benefit
in a grammar compiler
along the lines of YACC in the C world
or Happy~\cite{happy} and M{\'\i}mico~\cite{camarao01mimico}
in the Haskell world.
In addition to the parsing functions themselves,
the grammar compiler could automatically generate
the static ``derivations'' tuple type
and the top-level recursive ``tie-up'' function,
eliminating the problems of monadic representation
discussed in Section~\ref{monadic}.
The compiler could also reduce iterative notations
such as the popular `{\tt +}' and `{\tt *}' repetition operators
into a low-level grammar that uses only primitive constant-time operations,
preserving the linear parse time guarantee.
Finally, the compiler could rewrite left-recursive rules
to make it easier to express left-associative constructs in the grammar.

One practical area in which packrat parsing may have difficulty
and warrants further study
is in parsing interactive streams.
For example,
the ``read-eval-print'' loops in language interpreters
often expect the parser to detect at the end of each line
whether or not more input is needed to finish the current statement,
and this requirement violates the packrat algorithm's assumption
that the entire input stream is available up-front.
A similar open question is
under what conditions packrat parsing may be suitable
for parsing infinite streams.

\section{Conclusion}
\label{conclusion}

Packrat parsing is a simple and elegant method
of converting a backtracking recursive descent parser
implemented in a non-strict functional programming language
into a linear-time parser,
without giving up the power of unlimited lookahead.
The algorithm relies for its simplicity on
the ability of non-strict functional languages
to express recursive data structures with complex dependencies directly,
and it relies on lazy evaluation for its practical efficiency.
A packrat parser can recognize any language
that conventional deterministic linear-time algorithms can
and many that they can't,
providing better composition properties
and allowing lexical analysis to be integrated with parsing.
The primary limitations of the algorithm
are that it only supports deterministic parsing,
and its considerable (though asymptotically linear)
storage requirements.

\subsection*{Acknowledgments}

I wish to thank my advisor Frans Kaashoek,
my colleagues Chuck Blake and Russ Cox,
and the anonymous reviewers
for many helpful comments and suggestions.

\bibliography{parsing}
\bibliographystyle{plain}

\end{document}